\documentclass[twocolumn,prb,aps,longbibliography]{revtex4-1}
\usepackage{amsthm,amsfonts,graphicx,verbatim, color}
\usepackage{mathtools}
\usepackage{tabularx}
\usepackage{bm}
\usepackage[T1]{fontenc}
\usepackage[utf8]{inputenc}
\usepackage{physics}
\usepackage{braket}
\usepackage{multirow}
\usepackage{bigstrut}
\usepackage{booktabs}
\usepackage[colorlinks=true, linkcolor = blue, citecolor = blue, urlcolor = blue]{hyperref}

\graphicspath{{./figs/}{/Users/akshaykri/Princeton/Code/Fall_2019/Spin_chains/plots/}}

\begin{document}

\title{Beyond universal behavior in the one-dimensional chain with random nearest neighbor hopping }

\author{Akshay Krishna${}^1$ and R.\ N.\ Bhatt${}^{1,2}$}
\affiliation{${}^1$Department of Electrical Engineering, Princeton University, Princeton NJ 08544, USA \\
${}^2$School of Natural Sciences, Institute for Advanced Study, Princeton NJ 08540, USA}

\begin{abstract}
We study the one-dimensional nearest neighbor tight binding model of electrons with independently distributed random hopping and no on-site potential (i.e.\ off-diagonal disorder with particle-hole symmetry, leading to sub-lattice symmetry, for each realization).
For non-singular distributions of the hopping, it is known that the model exhibits a universal, singular behavior of the density of states $\rho(E) \sim 1/|E \ln^3|E||$ and of the localization length $\xi(E) \sim |\ln|E||$, near the band center $E = 0$.
(This singular behavior is also applicable to random XY and Heisenberg spin chains; it was first obtained by Dyson for a specific random harmonic oscillator chain).
Simultaneously, the state at $E = 0$ shows a universal, sub-exponential decay at large distances $\sim \exp [ -\sqrt{r/r_0} ]$.
In this study, we consider singular, but normalizable, distributions of hopping, whose behavior at small $t$ is of the form $\sim 1/ [t \ln^{\lambda+1}(1/t)  ]$, characterized by a single, continuously tunable parameter $\lambda > 0$.
We find, using a combination of analytic and numerical methods, that while the universal result applies for $\lambda > 2$, it no longer holds in the interval $0 < \lambda < 2$.
In particular, we find that the form of the density of states singularity is enhanced (relative to the Dyson result) in a continuous manner depending on the non-universal parameter $\lambda$; simultaneously, the localization length shows a less divergent form at low energies, and ceases to diverge below $\lambda = 1$.
For $\lambda < 2$, the fall-off of the $E = 0$ state at large distances also deviates from the universal result, and is of the form $\sim \exp  [-(r/r_0)^{1/\lambda}]$, which decays faster than an exponential for $\lambda < 1$.
\end{abstract}

\date{\today}
\maketitle


\section{Introduction \label{sec:intro}}

The study of transport in electronic systems in the non-interacting regime is a vast and thoroughly-researched area of physics.
The role of disorder in metal-insulator transitions is well-established\cite{Mott1961}, and the tendency of electrons to localize is understood in terms of universality classes based on dimensionality and symmetries\cite{GangofFour1979, Lee1985, Evers2008}.
In one dimension, the situation is relatively straightforward as quenched disorder generically causes Anderson localization\cite{Anderson1958} of the entire spectrum.
In this regime, transmission coefficients go to zero and all eigenstates are localized with exponential tails.
Over the years, several tight-binding models in 1-D lattices have been shown to circumvent this standard picture, usually by significant modification of the Hamiltonian, for example, by introducing correlations in the disorder\cite{Phillips1990, Moura1998, Izrailev1999, Moura1999}, adding long-range hopping\cite{Zhou2003}, or truncating the Hilbert space\cite{Krishna2018}.

For purely off-diagonal nearest-neighbor hopping disorder with no potential disorder, however, the model has been shown to have anomalous behavior near the center of the band ($E \to 0$).
This problem has a long history, dating back to the work of Dyson on the one-dimensional random harmonic oscillator chain with Poisson distributed couplings\cite{Dyson1953}.
Nearly two decades later, Smith showed that the problem mapped onto the nearest neighbor spin-1/2 XY chain with random couplings whose squares have a generalized Poisson distribution, and resulted in singular thermodynamic behavior in the $T \to 0$ limit\cite{Smith1970}.
Applying the Jordan-Wigner transformation\cite{Lieb1961} converts this to a spinless fermionic chain with nearest neighbor coupling, with singular density of states at the center of the band\cite{Theodorou1976}.
Eggarter and Riedinger showed that a universal result followed for any well-behaved distribution of hoppings\cite{Eggarter1978}.
As a result of several investigations, it is known that the density of states diverges as $\rho (E) \sim 1/|E \ln^3 |E||$\cite{Dhar1980}, and the localization length diverges as $\xi(E) \sim |\ln |E||$.
The state at zero energy is not conventionally extended, however, and decays with an envelope $\psi(r) \sim \exp \left( -\sqrt{r/r_0} \right)$ \cite{Fleishman1977, Soukoulis1981}.
The nature of the band center anomaly in this model is well documented in the literature\cite{Stone1981, Ziman1982, Evangelou1986, Roman1987, Markos1988, Roman1988, Inui1994}.

For brevity, we refer to the class of one-dimensional disordered nearest-neighbor hopping Hamiltonians with non-singular hopping distributions\footnote{Distributions of hopping $t$ that are less singular than the universal form $p(t) \sim 1/(t \ln^2(1/t))$ are also included} and the resultant singularity in $\rho (E)$ and $\xi(E)$ as the `Dyson' class and singularity respectively, in light of Freeman Dyson's pioneering work on the topic.

The magnetic version of this problem has historically attracted a lot of attention due to its experimental connection with the magnetic behavior of impurity-doped semiconductors\cite{Andres1981, Murayama1986, Sarachik1986, Paalanen1988} as well as organic chain complexes\cite{Bulaevskii1972, Tippie1981}.
The development of the strong disorder renormalization group (RG) technique\cite{Dasgupta1980, Bhatt1982, Fisher1992, Fisher1994, Fisher1995} systematically revealed the nature of the fixed point in this system, which controls the low-temperature thermodynamic behavior\cite{}.

Exact analytical\cite{McKenzie1996} and numerical\cite{Igloi1997} calculations have lent credence to the correctness of RG results.
Further studies have shed light on the nature of the Griffiths phases near the critical point\cite{Igloi1999, Igloi2001, Juhasz2006}.
Variations on this model, e.g., in the presence of multiple channels\cite{Brouwer2000}, 
and in higher dimensions\cite{Xiong2001}, have also been extensively researched.
The behavior of their densities of states\cite{Mudry2003, Rieder2014} and localization properties\cite{Brouwer2002, Gruzberg2005, Ryu2005} have been analyzed within the framework of universality.
It continues to be a topic of active research\cite{Mard2014, Zhao2015, Bera2016, Moure2018}.

The singularity is due to the sub-lattice symmetry of the Hamiltonian, which is technically a discrete unitary chiral symmetry, arising from combining particle-hole symmetry \emph{and} time-reversal symmetry on a bipartite lattice\cite{Chiu2016}.
It is known that additional terms in the Hamiltonian (e.g.\ on-site disorder, next-nearest-neighbor hopping or superconducting pairing\cite{Motrunich2001}) which lead to a breaking of the sub-lattice symmetry generically suppress the singular behavior.
Work to-date on the disordered nearest-neighbor hopping Hamiltonian has stayed within the class of hopping distributions that are non-singular, leading to the aforesaid universal behavior.
In this study, we remain within the nearest-neighbor hopping model, but consider the case of singular hopping distributions.

Our work is motivated by the observation that although the renormalized density of states is singular, it is \emph{not} the most singular form that is integrable.
In fact, it differs from such a singular function by a significant factor characterized by a \emph{finite} parameter.

As a consequence, in this work, we investigate the spectral and spatial properties of the standard nearest neighbor hopping Hamiltonian when the disordered hopping terms are drawn randomly from a probability distribution with a tunable sharp divergence at the origin of the form $1/[t \ln^{\lambda+1}(1/t))]$.
Such a distribution effectively converts the chain into islands of variable length, with very weak connections between the islands.
This regime was considered a corner case\cite{Fisher1994} and has eluded close scrutiny so far.
Nevertheless, as we show in this paper, for $\lambda < 2$ the low-energy behavior lies outside the standard universality class of Dyson-type models.
The Dyson singularity in the density of states is overwhelmed by the sharper singularity of the Hamiltonian itself.
The system shows a greater tendency to localize with the zero-energy wave function $\psi(r)$ asymptotically varying as $\exp \left[ -(r/r_0)^{\frac{1}{\lambda}} \right]$, which is stronger than the usual behavior.
At the boundary of the Dyson universality class ($\lambda=2$), we find $\psi(r) \sim \exp \left[ {-\sqrt{(r/r_0) \ln (r/r_0)}} \right]$.
Further, as $\lambda$ becomes smaller than 1, the distribution of wave function probability amplitudes becomes extremely broad, implying a rapid (super-exponential) decay of the envelope of the wave function.

We also throw light on a deep connection between this model and a completely different area of physics -- that of Brownian motion and diffusion.
The Schr\"{o}dinger equation of this model can be mapped to a random walk\cite{Eggarter1978}.
The universality of the Dyson class is akin to the universal nature of Fick's law of diffusion in which the mean squared displacement of a particle is proportional to the elapsed time i.e.\ $\langle x^2 \rangle \sim t$, \emph{irrespective} of the microscopic details of the diffusion process.
The non-universal results that we obtain here are parallel to the anolamous super-diffusion and L\'{e}vy flights seen when random walkers draw their step lengths from a heavy-tailed probability distribution.
In those cases, the displacement $\langle x^2 \rangle \sim t^\alpha$, with $\alpha>1$ a continuously tunable parameter\cite{Metzler2001, Zaburdaev2015}.
This is related to the non-universal and continuously tunable nature of singularity in the density of states and localization lengths in our model.

The rest of the paper is organized as follows.
In Sec.\ \ref{sec:hamiltonian}, we introduce the Hamiltonian and summarize our numerical methods.
Sec.\ \ref{sec:E0} contains results on the nature of the zero-energy eigenstate.
The density of states of our model and the localization length of the eigenstates away from the band center are discussed in Sec.\ \ref{sec:DoS} and \ref{sec:xi} respectively.
In these sections, we also elaborate on the connection to random walks.
In Sec.\ \ref{sec:exp}, we make some remarks on the connections to magnetic susceptibility in spin chains.
We conclude in Sec.\ \ref{sec:concl} with a summary of our results.


\section{Hamiltonian and numerical methods} \label{sec:hamiltonian}

We consider spinless electrons hopping on a one-dimensional chain with $N$ sites \begin{align}
H &= \sum\limits_{i=0}^{N-1} t_{i} \left( c_i^\dagger c_{i+1} + c_{i+1}^\dagger c_{i} \right), \label{eq:ham1}
\end{align}
where $c_i (c_i^\dagger)$ is the fermionic annihilation (creation) operator, and $t_{i}$ is the hopping matrix element between the $i$\textsuperscript{th} and $i+1$\textsuperscript{th} site.

As a consequence of sublattice symmetry, the spectrum of the system consists of pairs of eigenstates with energy $\pm E_m$.
The corresponding wave functions are $\Ket{\psi_m^\pm} = \frac{1}{\sqrt{2}} \left( \Ket{\psi_m^\text{e}} \pm  \Ket{\psi_m^\text{o}} \right)$, where $\Ket{\psi_m^\text{e}}$ and $\Ket{\psi_m^\text{o}}$ have weight only on the odd and even sites respectively. 
For notational simplicity and ease of computation, in the rest of the paper, we focus on the properties of the system for $E \ge 0$.
The situation at negative energies is identical by virtue of this symmetry.

The hopping terms $t_i$ are independent identically distributed random variables drawn from a normalized probability density
\begin{align}
p_\lambda(t) &= \begin{dcases}
\frac{c_\lambda}{t \ln^{\lambda+1} (d_\lambda/t) }, \quad &0 < t < e^{-(\lambda+1)} d_\lambda \\
0, &\text{otherwise.}
\end{dcases}\label{eq:p_lambda}
\end{align}
where $c_\lambda \equiv \lambda (\lambda+1)^\lambda$ is a normalization constant and $d_\lambda \equiv \exp \left( 2^\frac{1}{\lambda} (\lambda+1)\right)$ is a scaling factor to ensure that the median $t$ is unity.
The parameter $\lambda > 0$ controls the strength of the divergence at $t=0$, with $t \to \infty$ reducing $p(t)$ to a uniform distribution in $[0, 2]$.
For $\lambda > 2$, the system falls within the Dyson class.
However, for $0 < \lambda < 2$, we find non-universal scaling of the density of states and localization length.
This is the central finding of this paper.

For comparison, we show a family of probability distributions with a weaker power-law divergence,
\begin{align}
p_\alpha(t) &= \begin{dcases}
\frac{1-\alpha}{2 t^\alpha}, \quad &0 < t < 2^{\frac{1}{1-\alpha}}, \\
0, &\text{otherwise.}
\end{dcases}\label{eq:p_alpha}
\end{align}
where $0 \leq \alpha < 1$.
Here too, we normalize the distribution so that the median hopping is unity.
For $\alpha = 0$, we recover a uniform distribution in $[0, 2]$.
This distribution falls into the Dyson class for all values of $\alpha$, unlike the $p_\lambda$ distribution.

\begin{figure}[ht!]
\centering
\includegraphics[angle=0,width=\columnwidth]{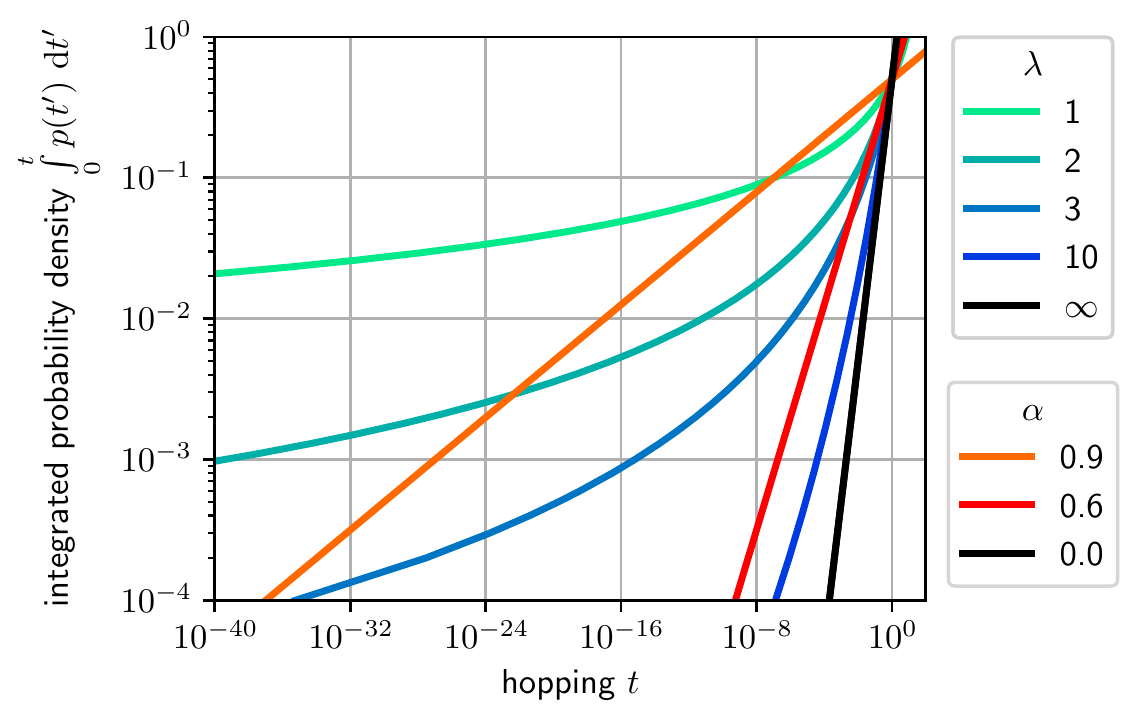} 
\caption{\label{fig:pdf} 
The integrated probability density of the hopping terms, on a logarithmic scale for both the $p_\lambda$ (Eq.\eqref{eq:p_lambda}), and $p_\alpha$ (power law, Eq.\ \eqref{eq:p_alpha}) distributions.
For $\alpha=0$ and $\lambda = \infty$, the distributions coincide and are equivalent to a box distribution.
It is evident that the $p_\lambda$ distributions have significant weight at exponentially small values of the hopping $t$, while the $p_\alpha$ distributions do not.}
\end{figure}

The contrast between the two families of probability densities can be gleaned from a plot of their cumulative distribution functions, Fig.\ \ref{fig:pdf}.
The $p_\lambda$ distributions are unusual in that they have a large weight at small $t$, so there is a significant probability of drawing exponentially small hopping terms $t$. 
We do not consider the $p_\alpha$ distribution in the rest of the paper.

Since we are primarily interested in the spectral distribution of energies and spatial behavior of wave functions, especially in the middle of the band, we perform strong disorder RG to extract the eigenvalues and eigenvectors.
We also perform transfer matrix calculations to extract the Lyapunov exponent as a measure of localization.
These are described below.

\subsection{Strong disorder renormalization group}

We perform a 4-site RG on a system with periodic boundary conditions, starting with the strongest hopping, following the standard method\cite{Mard2014, Johri2014}. 
At each iteration, the system size reduces by two sites, in a manner that respects the sublattice symmetry of the system.

The RG proceeds by identifying the neighboring sites with the largest absolute value of hopping in the system.
Let this value be $t_{i}$.
The four sites considered are $\ket{i-1}, \ket{i}, \ket{i+1}$ and $\ket{i+2}$, with hopping terms $t_{i-1}, t_{i}$ and $t_{i+1}$. 
The RG step assumes that $t_{i} \gg t_{i-1}, t_{i+1}$ and computes a second-order perturbation theory solution for this 4-site subsystem.
In practice, the RG scheme works reasonably well even when this condition is not met.
Its accuracy increases dramatically as the iterations proceed and the energy scale becomes lower, and is therefore reliable for the states at the center of the band that are of interest to us.

The RG step updates the two outer states as follows \begin{align*}
\ket{i-1} &\to \ket{i-1} - \frac{t_{i-1}}{t_{i}} \ket{i+1},\\
\ket{i+2} &\to \ket{i+2} - \frac{t_{i+1}}{t_{i}} \ket{i}.
\end{align*}
It also generates a new hopping of strength $-\frac{t_{i-1} t_{i+1}}{t_{i}}$ between them.

The middle two states hybridize with each other as \begin{align*}
\ket{+} = \ket{i} + \ket{i+1} + \frac{t_{i-1}}{t_{i}} \ket{i-1} + \frac{t_{i+1}}{t_{i}} \ket{i+2},\\
\ket{-} = \ket{i} - \ket{i+1} - \frac{t_{i-1}}{t_{i}} \ket{i-1} + \frac{t_{i+1}}{t_{i}} \ket{i+2}.
\end{align*}
These states have energies $E_+ = t_{i} + \frac{t_{i-1}^2 + t_{i+1}^2}{2 t_{i}}$ and $E_- = -t_{i} - \frac{t_{i-1}^2 + t_{i+1}^2}{2 t_{i}}$ respectively, and are removed from the system.

The decimation proceeds in this manner by successively identifying the bond with the largest hopping, until there are only two sites left in the system.
At that point, the $2 \times 2$ matrix for these two sites is solved exactly to obtain the last two eigenstates.  

\subsection{Transfer matrix method}

The transfer matrix method is an efficient method of solving for the wave function $\Ket{\psi}$ of a one-dimensional lattice model at fixed energy $E$ with open boundary conditions\cite{MacKinnon1993}.
It follows directly from Schr\"{o}dinger's equation for the Hamiltonian (Eq.\ \eqref{eq:ham1}) and relates the probability amplitudes $\psi_i \equiv \Braket{i|\psi}$ between successive pairs of sites as 
\begin{align}
\left( \begin{array}{c}
\psi_{i+1}\\
\psi_i 
\end{array} \right) &= T_i(E) \left( \begin{array}{c}
\psi_{i}\\
\psi_{i-1} 
\end{array} \right) \nonumber \\ 
\text{where } T_i(E) &= \left( \begin{array}{cc}
\frac{E}{t_{i}} & -\frac{t_{i-1}}{t_{i}} \\
1 & 0
\end{array} \right) \label{eq:TM}
\end{align}
In the limit $N \to \infty$, the behavior of the (unnormalized) wave function obtained by iterating this equation is correct and independent of the initial conditions $\psi_0$ and $\psi_1$.

The product of transfer matrices over $N$ sites is $Q_N(E) \equiv T_N(E) \times \cdots \times T_1(E)$.
We mention here that the Lyapunov exponent $\gamma(E)$ is found from the eigenvalues of this product matrix as \begin{align}
\gamma(E) \equiv \ln \max \text{eig} \left( Q_N(E)^\dagger Q_N(E) \right)^{\frac{1}{2N}}.
\end{align}
The Lyapunov exponent is commonly taken as a proxy for the inverse localization length of the state.
However, anticipating the non-exponential decay of wave functions in our model, we use a slightly different metric to quantify the localization, as described in the next section.


\section{The band center state} \label{sec:E0}

From the transfer matrix equation Eq.\ \eqref{eq:TM}, \begin{align}
t_{i} \psi_{i+1} + t_{i-1} \psi_{i-1} = E \psi_i. \label{eq:ham2}
\end{align}
For a system with an odd number of sites $N$ with open boundary conditions, there exists a solution to Eq.\ \eqref{eq:ham2}, with $E=0$, such that odd-numbered sites carry no weight $(\psi_{2i+1} = 0)$ and even-numbered sites are such that \begin{align}
\psi_{2i} &= (-1)^i \frac{\prod\limits_{j=0}^{i-1} t_{2j}}{\prod\limits_{j=0}^{i-1} t_{2j + 1}}\psi_0. \label{eq:ham2a}
\end{align}

To determine the envelope of this wave function, similar to the approach of Fleishman and Licciardello\cite{Fleishman1977}, we write \begin{align}
\ln \left\lvert \frac{\psi_{2n}}{\psi_0} \right\rvert &= \sum\limits_{i=0}^{n-1} u_{2i+1} - \sum\limits_{i=0}^{n-1} u_{2i}, \label{eq:Flei1}
\end{align}
where the distribution of $u_i \equiv - \ln t_i$ is found by transforming Eq.\ \eqref{eq:p_lambda}.
We obtain  \begin{align}
p_\lambda(u) &= \frac{c_\lambda}{(u + \ln d_\lambda)^{\lambda+1}}, \quad \lambda+1 - \ln d_\lambda \leq u < \infty.
\end{align}
The coefficients $c_\lambda$ and $d_\lambda$ above are the same as in Eq.\ \eqref{eq:p_lambda}.

For $\lambda > 2$, the $u_i$'s have a finite mean $\mu_u$ and variance $\sigma_u^2$, and applying the Central Limit Theorem (CLT) to Eq.\ \eqref{eq:Flei1} leads to the established result that \begin{align}
\frac{1}{\sqrt{2n}} \ln \left\lvert \frac{\psi_{2n}}{\psi_0} \right\rvert \xrightarrow{n \to \infty} \mathcal{N}(0, \sigma_u^2). \label{eq:CLT1}
\end{align}
Here $\mathcal{N}(\mu, \sigma^2)$ denotes a normal distribution with mean $\mu$ and standard deviation $\sigma$.

If $\lambda \leq 2$, then the $u_i$'s in Eq.\ \eqref{eq:Flei1} have infinite variance, and the standard (Gaussian) CLT does not apply.
We appeal to the Generalized CLT\cite{Gnedenko1968, Shintani2018}, which specifies limiting distributions for sums of random variables with heavy power-law tails.
These are known as L\'{e}vy alpha-stable distributions, denoted as $\mathcal{S}(\lambda, \beta, \gamma, \delta)$ in the literature, with $\lambda \in (0, 2]$, $\beta \in [-1, 1]$, $\gamma \in (0, \infty)$ and $\delta \in (-\infty, \infty)$ representing stability, skewness, scale and shift parameters respectively.
As a result, Eq.\ \eqref{eq:CLT1} is modified to
\begin{align}
\frac{1}{(2n)^{1/\lambda}} \ln \left\lvert \frac{\psi_{2n}}{\psi_0} \right\rvert &\xrightarrow{n \to \infty} \mathcal{S}\left( \lambda, 0, \gamma(\lambda), 0 \right), \label{eq:CLT2} \\
\text{where } \gamma(\lambda) &= \left( \frac{\pi c_\lambda}{2\lambda \sin(\frac{\pi \lambda}{2}) \Gamma(\lambda)} \right)^{\frac{1}{\lambda}} \label{eq:gamma}.
\end{align}
Here $\mathcal{S}(\lambda, 0, \gamma, 0)$ is a \emph{symmetric} L\'{e}vy  alpha-stable distribution. 
The probability density functions $p(x)$ of these distributions are not expressible in closed form, except for some special cases.
They have tails of the form $p(x) \sim 1/|x|^{\lambda+1}$ as $x \to \pm \infty$, and are usually specified in terms of their characteristic function \begin{align}
\tilde{p}(k) = \langle e^{ikx} \rangle = e^{-|\gamma k|^\lambda}.
\end{align}
Two well-known cases are when the stability parameter $\lambda = 1$, for which it reduces to a Lorentzian, and when $\lambda=2$, which corresponds to a Gaussian.

For the marginal case $\lambda=2$, the scaling law of the generalized CLT leads to \begin{align}
\frac{1}{\sqrt{2n \ln n}} \ln \left\lvert \frac{\psi_{2n}}{\psi_0} \right\rvert \xrightarrow{n \to \infty} \mathcal{N} \left(0, \frac{c_2}{2}\right). \label{eq:CLT3}
\end{align}

Combining all three cases, we find that a wave function decay length scale $r_0$ may be defined as \begin{align}
\frac{1}{r_0} &\equiv \lim\limits_{n \to \infty} \frac{1}{f(2n)} \left\lvert \ln \left\lvert \frac{\psi_{2n}}{\psi_0} \right\rvert \right\rvert, \text{ with } \\
f(n) &= \begin{dcases}
\sqrt{n}, &\lambda > 2\\
\sqrt{n \ln n}, &\lambda = 2\\
n^{\frac{1}{\lambda}}, & 0<\lambda<2,
\end{dcases} \label{eq:fr}
\end{align}
such that $r_0$ has a well-normalized and non-singular (but possibly heavy-tailed) probability distribution, and hence the wave function has fluctuations of the form $\psi(r) \sim e^{-f(r/r_0)}$.

Having thus established the functional form of the envelope of the wave function in Eq.\ \eqref{eq:fr} for the full parameter range $0 < \lambda < \infty$, we turn our attention to calculating the distribution of the inverse length $1/r_0$.
From Eq.\ \eqref{eq:CLT1}, the distribution of $1/r_0$ for the Dyson class ($\lambda \geq 2$) is a one-sided Gaussian.
The Gaussian is one-sided (i.e.\ no support for negative values) as we define $1/r_0 > 0$ to be the length scale of a \emph{decaying}, and not growing wave function.
For $\lambda < 2$ then, the distribution of $1/r_0$ is also one-sided, obtained by folding the symmetric L\'{e}vy  alpha-stable distribution $\mathcal{S}\left( \lambda, 0, \gamma(\lambda), 0 \right)$ at the origin.
Since these distributions are extremely broad (and for $\lambda \leq 1$ have infinite mean), we find it convenient to characterize them by their median instead, which is finite for all $\lambda$.
The median, denoted $\bar{r}_0^{-1}$, represents a typical value of this quantity.

Analytic expressions for $\bar{r}_0^{-1}$ are obtained as follows.
From the CLT, Eq.\ \eqref{eq:CLT1}, we predict that for $\lambda > 2$, $\bar{r}_0^{-1}(\lambda) = \sqrt{2} \ \text{erfi} \left(\frac{1}{2}\right) \sigma_u(\lambda)$, with the variance of the logarithm of the hoppings \begin{align}
\sigma^2_u(\lambda) = \frac{\lambda (\lambda+1)^2}{(\lambda-2)(\lambda-1)^2}, \label{eq:sigma_u}
\end{align}
and the inverse error function $\text{erfi}(x)$ arising from integrating the Gaussian distribution.
At $\lambda = 2$, from Eq.\ \eqref{eq:CLT3}, $\bar{r}_0^{-1}(\lambda) = 3 \sqrt{2} \ \text{erfi} \left(\frac{1}{2}\right) \approx 2.02$.
For strongly divergent hopping distributions with $\lambda < 2$, we find that $\bar{r}_0^{-1}(\lambda) = \gamma(\lambda) Q_\lambda \left( \frac{3}{4} \right)$, where $\gamma(\lambda)$ is the scale factor in Eq.\ \eqref{eq:gamma}.
$Q_\lambda(f)$ denotes the quantile function, obtained by inverting the cumulative density function, such that $\int\limits_{-\infty}^{Q_\lambda(f)} p(x) \mathrm{d} x = f$.
Here $p(x)$ is the probability density corresponding to a unit-scaled alpha-stable random variable $\mathcal{S}(\lambda, 0, 1, 0)$.

We verify the theoretical predictions above using numerical techniques, similar to the approach of Inui et al\cite{Inui1994}.
There, they define a correlation function $\left\langle \left\lvert \ln \left\lvert \psi_{r+j} \right\rvert - \ln \left\lvert \psi_j \right\rvert \right\rvert \right\rangle$, where the angular brackets denote the mean over all sites $j$ as well as realizations of disorder.
We generalize their correlation function by defining it in terms of the median over all sites and realizations of disorder as \begin{align}
g(r) \equiv \text{median} \left\lvert \ln \left\lvert \frac{\psi_{r+j}}{\psi_j} \right\rvert \right\rvert. \label{eq:gr}
\end{align}
Then we compute $\bar{r}_0^{-1}$ as \begin{align}
\bar{r}_0^{-1} \equiv \lim\limits_{r \to \infty} \frac{g(r)}{f(r)},
\end{align}
with $f(r)$ as in Eq.\ \eqref{eq:fr}.

\begin{figure}[ht!]
\centering
\includegraphics[angle=0,width=\columnwidth]{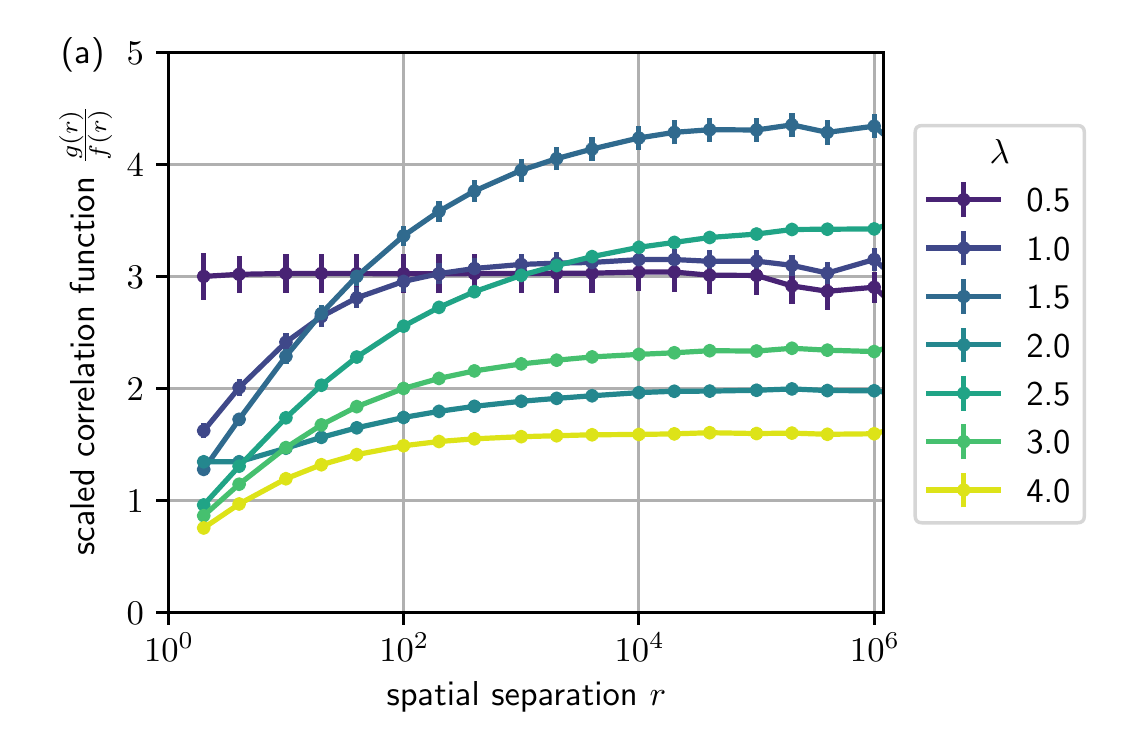} 
\includegraphics[angle=0,width=\columnwidth]{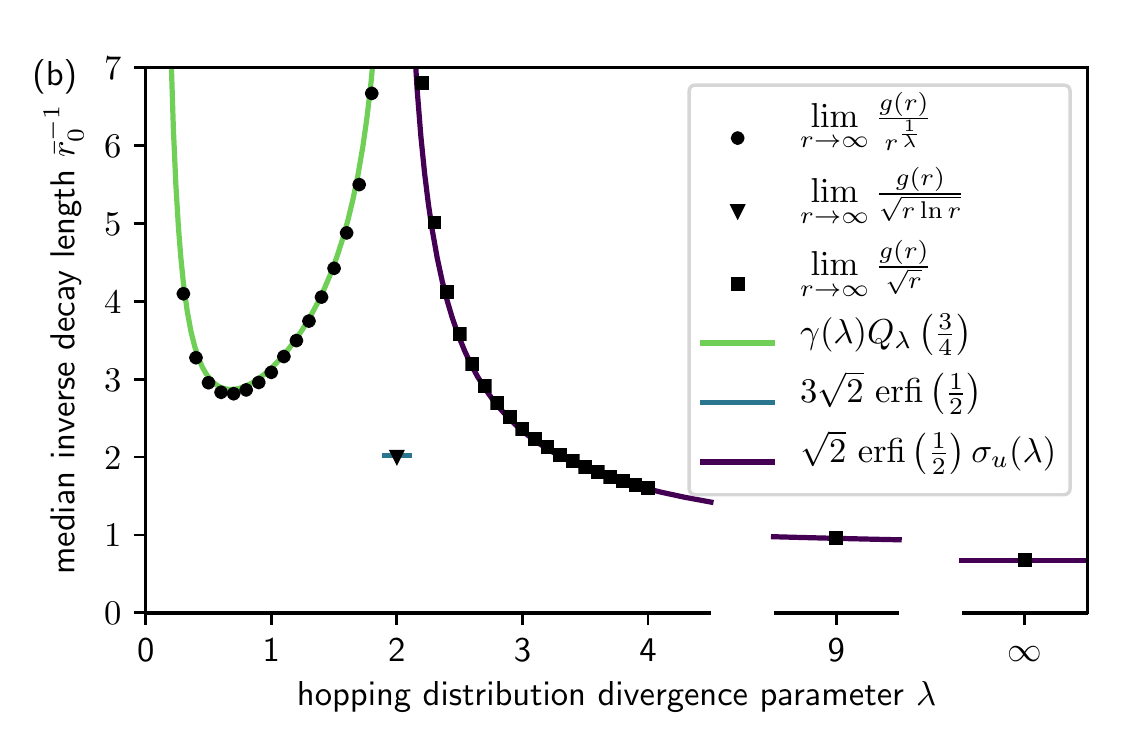} 
\caption{\label{fig:zeromode} (a) The scaled correlation function $g(r)/f(r)$ (see Eqs.\ \eqref{eq:fr} and \eqref{eq:gr}) for systems of size $N = 20\ 000\ 001$.
Ensemble medians are calculated over the middle $10^7$ sites of the system for the $E=0$ state for 100 realizations of disorder. 
Several some values of the divergence parameter $\lambda$ are shown.
(b) The numerical and predicted theoretical values of the median inverse decay length $\bar{r}_0^{-1}$.
Numerical values (black markers) are obtained by fitting a function of the form $\frac{g(r)}{f(r)} = 1/\bar{r}_0 + b/(\ln r)^c$ to the curves in panel (a) for $10^3 \leq r \leq 10^6$.
See text for a description of theoretical results (solid lines).}
\end{figure}

We compute the zero-energy state for large systems using Eq.\ \eqref{eq:ham2a}.
In Fig.\ \ref{fig:zeromode}(a), we plot the scaled correlation function $g(r)/f(r)$ thus obtained for various values of the hopping probability divergence parameter $\lambda$.
We find that as $r \to \infty$, the scaled correlation function tends to a constant, as expected.
Hence, from the asymptote at large $r$, the typical inverse decay length $\bar{r}_0^{-1}$ is extracted.

In Fig.\ \ref{fig:zeromode}(b), we find that numerically obtained decay length agrees closely with our analytical results in all three regimes, demonstrating the non-universal $e^{-f(r/r_0)}$ decay of the wave function.

In the next section, we explore the behavior of the system away from zero energy.


\section{Density of States} \label{sec:DoS}

The asymptotic form of the low-energy density of states of the Dyson model is controlled by the infinite randomness fixed point, to which all `well-behaved' initial distributions of disorder flow under the RG.
This is
\begin{align}
\rho(E) \simeq \frac{\sigma_u^2}{E |\ln E|^3} \label{eq:Egg1},
\end{align}
if the hopping distribution is well-behaved.
In this case, the integrated density of states $N(E)$ is \begin{align}
N(E) &\equiv \int\limits_{-E}^E \mathrm{d}E' \rho (E') = \frac{\sigma_u^2}{|\ln E|^2} \label{eq:Egg2}.
\end{align}

However, the singular distribution Eq.\ \eqref{eq:p_lambda} lies outside the basin of attraction of this fixed point, if $\lambda \leq 2$, because $\sigma_u^2 \to \infty$.
One might then conjecture that for the non-Dyson case, Eq.\ \eqref{eq:Egg2} for the density of states must be generalized to \begin{align}
N(E) = \frac{b_\lambda}{|\ln E|^{\lambda}},
\end{align}
where $b_\lambda$ is a $\lambda$-dependent parameter.
One might further expect that the density of states is given by the distribution of bare hoppings $p_\lambda(t)$ itself, because the singularity in $p_\lambda(t)$ is so sharp that it is not affected by the decimation procedure.
This would imply that \begin{align}
b_\lambda = c_\lambda/\lambda = (\lambda+1)^\lambda. \label{eq:naive}
\end{align}

However, this guess is only partially correct.
The right result is obtained by mapping this system to a random walk, which we describe below.
Such a mapping was first described by Eggarter and Riedinger\cite{Eggarter1978} for the Dyson class, so we recap their work first, and then extend their results to the non-Dyson case.

Corresponding to the on-site wave function amplitudes $\psi_i$, one defines `self-energies' \begin{align}
\Delta_i \equiv t_{i-1} \frac{\psi_{i-1}}{\psi_i}.
\end{align}
In terms of these self-energies, the Schr\"{o}dinger equation, Eq.\ \eqref{eq:ham2}, is re-written as \begin{align}
\Delta_{i+1} &= \frac{t_i^2}{E - \Delta_i}, \label{eq:SE1} \\
\Delta_{i+2} &= \left( \frac{t_{i+1}}{t_i} \right)^2 \Delta_i \left[ \frac{1-\frac{E}{\Delta_i}}{1 + \frac{E \Delta_i}{t_i^2}(1-\frac{E}{\Delta_i})} \right]. \label{eq:SE2}
\end{align}
The integrated density of states is related to the fraction $f_+$ of positive self-energies in the sequence $\{ \Delta_i \}$ as\cite{Schmidt1957} \begin{align}
N(E) = 2f_+ -1.
\end{align}
When $E = 0$, the term in the square brackets of Eq.\ \eqref{eq:SE2} may be ignored and the signs of the self-energies $\Delta_i$ alternate.
Since there are as many positive self-energies as there are negative self-energies, the integrated density of states $N(0) = 0$, as expected.

In this regime, Eq.\ \eqref{eq:SE2} is equivalent to a discrete-time random walk \begin{align}
\ln \Delta_{i+2} = 2 \left( u_{i+1} - u_i \right) + \ln \Delta_i,
\end{align}
where $u_i \equiv \ln t_i$ is a random increment as before.
With the mapping $(\ln \Delta, i) \mapsto (x, t)$, the connection to the continuum Langevin equation becomes transparent \begin{align}
\frac{\mathrm{d}x}{\mathrm{d}t} = W_\lambda \zeta_\lambda(t), \label{eq:Langevin}
\end{align}
where a particle's position as a function of time $x(t)$ is governed by a unit-strength delta-correlated noise process $\zeta_\lambda(t)$ with scale factor $W_\lambda$.

When $\lambda > 2$, $\zeta_\lambda(t)$ is Gaussian white noise, and the stochastic process above is Brownian motion. 
But when $\lambda < 2$, $\zeta_\lambda(t)$ is a symmetric white alpha-stable noise\cite{Janicki1994}.
Such a stochastic process is known as a L\'{e}vy flight\cite{Klages2008}, and exhibits remarkably different behavior from a regular diffusive random walk.
In a L\'{e}vy flight, the stochastic process Eq.\ \eqref{eq:Langevin} may be interpreted as the limiting case of the discrete random walk when the time increments $\Delta t \to 0$ \begin{align}
\Delta x = x(t + \Delta t) - x(t) = W_\lambda (\Delta t)^{1/\lambda} \zeta_\lambda,
\end{align}
where $\zeta_\lambda$ is a random variable drawn from the symmetric alpha-stable distribution $\mathcal{S}(\lambda, 0, 1, 0)$.
The scale factor $W_\lambda$ is found by applying the (generalized) CLT as in the previous section, it is \begin{align}
W_\lambda = \begin{dcases}
2 \sigma_u(\lambda), \quad &\lambda > 2 \quad (\text{Brownian motion}),\\
2 \gamma(\lambda), \quad &\lambda < 2 \quad (\text{L\'{e}vy flight}),
\end{dcases}
\end{align}
where $\sigma_u(\lambda)$ and $\gamma(\lambda)$ are the same as in Eq.\ \eqref{eq:sigma_u} and \eqref{eq:gamma} respectively.

When $E$ is non-zero, but small and positive, this random-walk picture still holds as long as the self-energies are in the regime \begin{align}
E \ll \Delta_i \ll \tilde{t}^2/E, \label{eq:endpoints}
\end{align}
so that the term in the square brackets of Eq.\ \eqref{eq:SE2} is small.
Here $\tilde{t}$ is a typical value of $t$.
The constraints enforced by Eq.\ \eqref{eq:endpoints} manifest themselves in boundary conditions -- when $\Delta_i$ is comparable to or larger than $\tilde{t}^2/E$, the denominator in the square brackets ensures that $\Delta_i$ does not increase any further, and when $\Delta_i$ is smaller than $E$, then by Eq.\ \eqref{eq:SE1}, the alternating sequence of the sign of the $\Delta_i$'s is broken by two consecutive positive values.
The random-walk then starts afresh from this site, at a value $\Delta_{i+1} \simeq \tilde{t}^2/E$.

So when $E \neq 0$, instead of a particle executing free Brownian motion or a free L\'{e}vy flight, we have a particle confined along a finite portion of the x-axis.
There is an infinite potential barrier at $x = \ln (\tilde{t}^2/E)$ and an absorbing barrier at $x = \ln E$.
The particle originates just below $x = \ln (\tilde{t}^2/E)$, and the time it spends above $x = \ln E$ before being absorbed is the length of the self-energy sequence with alternating sign.
The fraction of excess positive self-energies $(2 f_+ -1)$, and hence the integrated density of states, is inversely proportional to the mean length of this sequence.

This implies, in terms of the so-called mean first-passage time $T_{\text{FP}}$ of the problem, the desired integrated density of states is simply \begin{align}
N(E) = \frac{1}{T_{\text{FP}}(\lambda)}.
\end{align}
The quantity $T_{\text{FP}}(\lambda)$ is the answer to the question -- how long, on average, does it take a particle to first reach the position $x_f = \ln E$, after being released at initial position $x_0 = \ln (\tilde{t}^2/E)$, and under the influence of a stochastic process Eq.\ \eqref{eq:Langevin} with an infinite potential barrier $V(x) = \infty$ for $x > \ln \tilde{t}^2/E$?
The scale-invariance of this process implies that \begin{align}
T_{\text{FP}}(\lambda) &= \begin{dcases}
\tau_\lambda \left( \frac{L}{W_\lambda} \right)^2, \ & \lambda > 2 \enskip (\text{Brownian motion}), \\
\tau_\lambda \left( \frac{L}{W_\lambda} \right)^\lambda, \ & \lambda < 2 \enskip (\text{L\'{e}vy flight})
\end{dcases}
\end{align}
where $L = x_f - x_0 = 2 \ln (\tilde{t}/E)$ is the length of the random walk, and $\tau_\lambda$ is the mean first passage time for the stochastic process Eq.\ \eqref{eq:Langevin} with unit strength over a unit interval, and is a number of order one.
For the Gaussian process in the case of Brownian motion, $\tau_\lambda = 1$ by standard techniques, e.g.\ the method of images\cite{Balakrishnan2008}.
First passage times in L\'{e}vy flights have been the subject of intense exploration in recent years\cite{Koren2007, Dybiec2017, Metzler2019, Palyulin2019}.
However, to the best of our knowledge there are no analytical results for $\tau_\lambda$: the case of a L\'{e}vy flight on finite 1-D domain with the particle being released adjacent to the reflecting boundary.
Nevertheless, it is fairly straightforward to compute $\tau_\lambda$ by numerical simulations (see Appendix \ref{apptau}), and in terms of this quantity, the integrated density of states is given by
\begin{align}
N(E) = \begin{dcases}
\frac{\sigma_u^2}{|\ln E|^2}, \ & \lambda > 2, \\
\frac{[\gamma(\lambda)]^\lambda}{\tau_\lambda |\ln E|^\lambda}, \ & \lambda < 2.
\end{dcases} \label{eq:DoS_th}
\end{align}

We are left with the boundary case $\lambda = 2$.
In the limit $\lambda \to 2$, both expressions above have a pre-factor that goes to infinity, but the same functional dependence on $\ln E$.
Inspired by the logarithmic behavior of the envelope of the zero-energy state at the boundary of the Dyson class described in the previous section, we conjecture that the density of states is \begin{align}
N(E) &= \frac{b_2 \ln |\ln E|}{|\ln E|^2}, \ & \lambda = 2 \enskip. \label{eq:DoS_th2}
\end{align}

\begin{figure}[ht!]
\centering
\includegraphics[angle=0,width=\columnwidth]{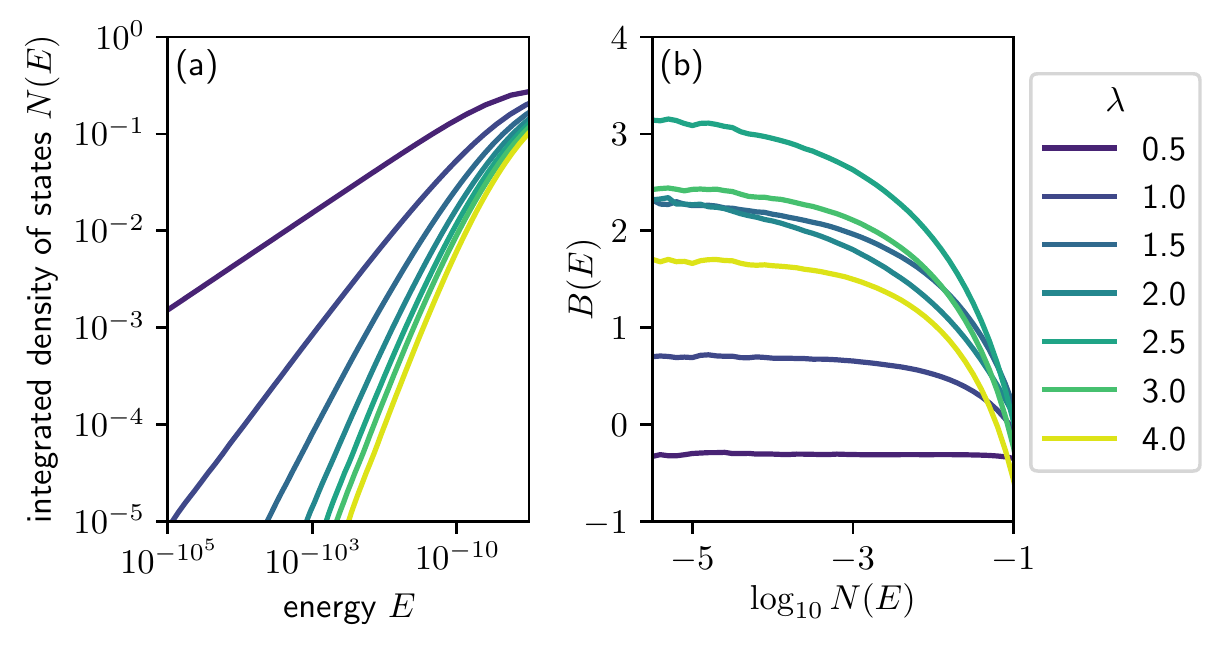}
\includegraphics[angle=0,width=\columnwidth]{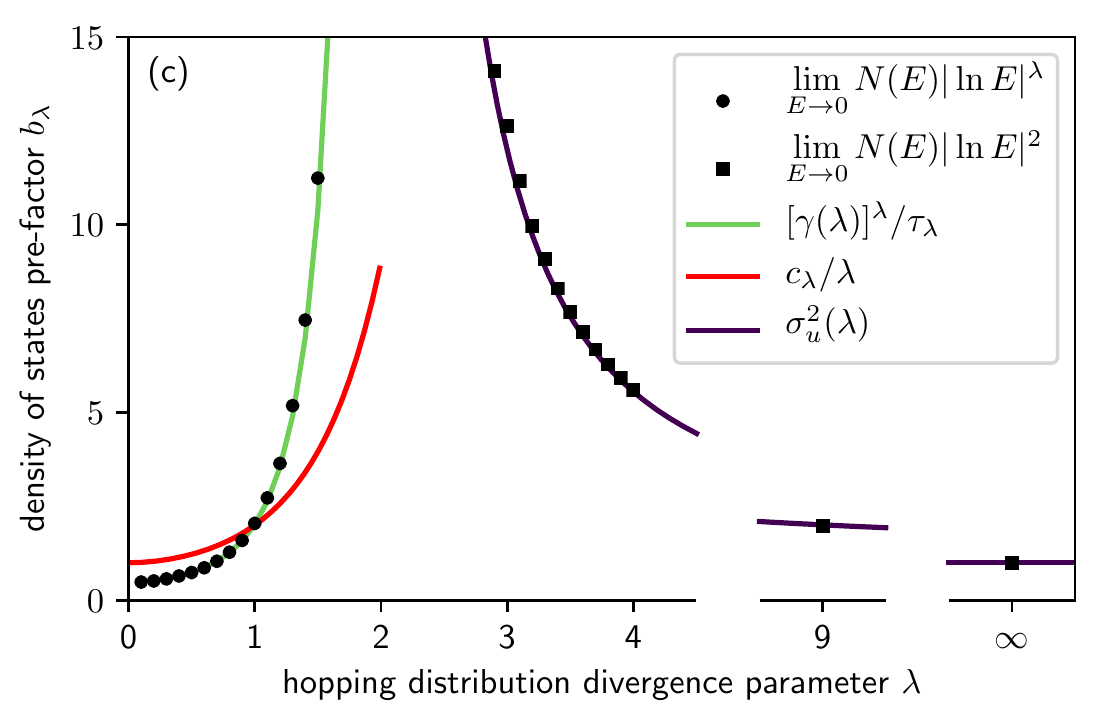}  
\caption{\label{fig:DoS1} 
(a) The integrated density of states $N(E)$, obtained from numerical strong disorder RG calculations, for different values of $\lambda$, .
The straight line behavior with slope $\min(\lambda, 2)$ is clearly seen.
Note the scaling on the horizontal axis is a logarithm of a logarithm.
Not all values of $\lambda$ are shown.
(b) To extract the functional form of $N(E)$ at low energies, we plot $B(E)$ (see Eq.\ \eqref{eq:BE}) by the appropriate transformations on the curves in (a).
(c) The density of states pre-factor $b_\lambda$ (black markers) obtained by a fitting the curves in (b) to a function of the form $B(E) = b_\lambda + c/(\ln N(E))^d$.
Green and purple lines are the theoretical predictions in Eq.\ \eqref{eq:DoS_th}.
The red line is the naive prediction, Eq.\ \eqref{eq:naive}.
}
\end{figure}

We validate the predictions in Eqs.\ \eqref{eq:DoS_th} and \eqref{eq:DoS_th2} numerically.
In Fig.\ \ref{fig:DoS1}(a), we plot the integrated density of states at the low-energy end of the spectrum for different $\lambda$.
At each value of $\lambda$, data are collected by running the strong disorder RG procedure on 200 realizations of disorder on systems of size $N = 10^6$ sites, for a total of $10^8$ unique eigenvalues. 
This enables us to study the extreme low-energy tail of the spectrum, upto $E \lesssim 10^{-10^6}$.

In order to test the power laws and obtain the pre-factors in Eq.\ \eqref{eq:DoS_th}, we compute and plot the function $B(E)$ in Fig.\ \ref{fig:DoS1}(b).
\begin{align}
B(E) &= \begin{dcases}
\ln N(E) + 2 \ln |\ln E|, &\lambda > 2, \\
\ln N(E) + 2 \ln |\ln E| - \ln (\ln |\ln E|), &\lambda = 2, \\
\ln N(E) + \lambda \ln |\ln E|, &\lambda < 2.
\end{dcases} \label{eq:BE}
\end{align}
It is apparent that $B(E)$ tends to a constant at small energies for all $\lambda$, thus corroborating the dependence of $N(E)$ on $\ln E$ predicted in Eqs.\ \eqref{eq:DoS_th} and \eqref{eq:DoS_th2}.

In Fig.\ \ref{fig:DoS1}(c), we plot the numerical value of the pre-factor extracted from fitting the curves in Fig.\ \ref{fig:DoS1}(b).
We find that in the both the Dyson class ($\lambda > 2$) as well as beyond the Dyson class ($\lambda < 2$), there is a clear agreement between the analytical prediction from the mapping to random walks, and our numerical fit.
The unknown value $b_2$ in Eq.\ \eqref{eq:DoS_th2} is found to be $18 \pm 2$.
The naive assumption (Eq.\ \eqref{eq:naive}) that the decimation procedure does not affect the density of states for $\lambda < 2$ is not correct.

It is worth pointing out that for $\lambda < 1$, we find that $[\gamma(\lambda)]^\lambda / \tau_\lambda < c_\lambda / \lambda$, so the density of states pre-factor is \emph{smaller} than that of the probability density of hoppings.
This means the strong disorder RG procedure causes the density of states singularity at $E=0$ to retain the functional form given by the distribution of bare hoppings, but its magnitude is weakened.
This is in contrast to the usual state of affairs, where the renormalization always broadens the distribution of couplings, and enhances the size of the singularity.


\section{Localization lengths} \label{sec:xi}

\begin{figure*}[ht!]
\centering
\includegraphics[angle=0,width=\textwidth]{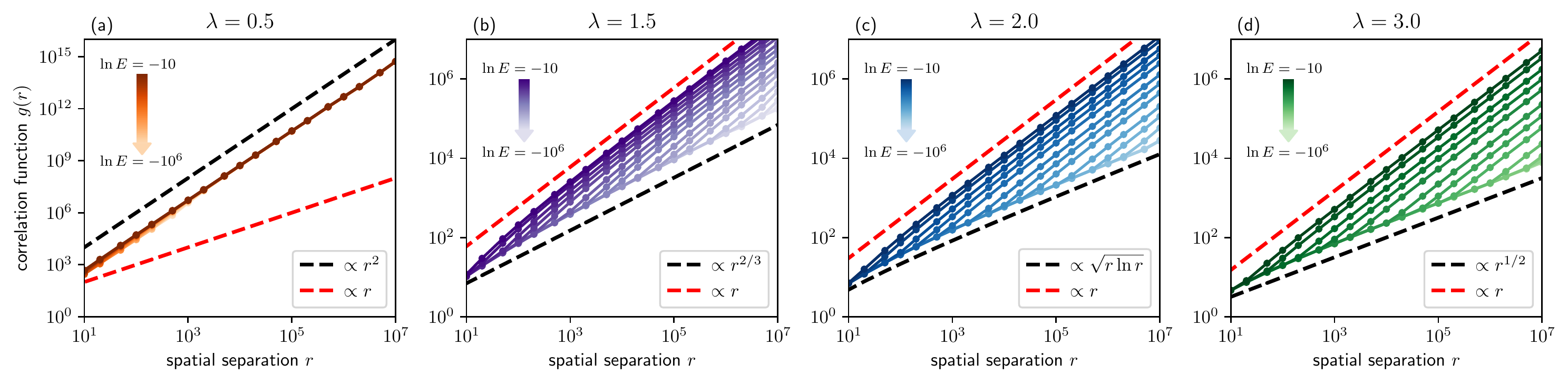} 
\caption{\label{fig:xi_scaling2}
The correlation function $g(r)$ (Eq.\ \eqref{eq:gr}) for a range of non-zero energies ($E = e^{-10}$ to $E = e^{-10^6}$) for four representative values of $\lambda$.
Each point is obtained by considering the middle $10^7$ sites of 100 eigenstates, each of size $N = 2 \times 10^7$.
Darker shades denote higher energies, and shades become lighter as $E \to 0$.
Dashed red and black lines are drawn as guides to the eye.
At low energies and short length scales, the scaling of the correlation function is dominated by that of the critical point ($E=0$), so that when $\lambda < 2$ (panels (a) and (b)), $g(r) \sim r^{1/\lambda}$.
When $\lambda = 2$, $g(r) \sim \sqrt{r \ln r}$ (panel (c)), and when $\lambda > 2$, $g(r) \sim \sqrt{r}$ as in the Dyson class.
At any fixed energy, the correlation function for $\lambda > 1$ shows a kink as at a particular value of $r$, beyond which it shows the trademark linear scaling characteristic of exponential localization (panels (b)-(d)).
When $\lambda < 1$ (panel (a)), the entire spectrum is super-exponentially localized, so that $g(r) \sim r^{1/\lambda}$ irrespective of energy.
}
\end{figure*}

For states with an exponentially decaying envelope, the localization length $\xi$ quantifies how fast the wave function falls off: $\psi(r) \sim \exp(-r/\xi)$.
This typically depends on the energy $E$.
The Thouless relation\cite{Thouless1972} connects the density of states with the localization length \begin{align}
\frac{1}{\xi(E)} &= \int\limits_{-\infty}^{\infty} \mathrm{d}E' \  \rho(E') \ln |E'-E| + \mu_u, \label{eq:Thou1}
\end{align}
where $\mu_u \equiv \langle - \ln t \rangle$ is the mean logarithm of the reciprocal of the hopping terms.
This relationship is applicable whenever $\mu_u$ is finite, i.e.\ for the Dyson class ($\lambda > 2$) as well as the non-universal case when $1 < \lambda \leq 2$.
In these cases, the state at zero energy is sub-exponentially localized, so $\xi(0) \to \infty$, and we obtain the sum rule \begin{align}
\int\limits_{0}^{\infty} \mathrm{d}E \  \rho(E) \ln(E) = -\frac{\mu_u}{2} \label{eq:sumrule}.
\end{align}
As shown by Theodorou and Cohen\cite{Theodorou1976}, one can combine Eqs.\ \eqref{eq:Thou1} and \eqref{eq:sumrule}, to obtain the leading behavior of the localization length at non-zero energy.
 \begin{align}
\frac{1}{\xi(E)} &= 2 \int\limits_{0}^{E} \mathrm{d}E' \  \rho(E') \ln \left( \frac{E}{E'} \right) = \int\limits_0^E \mathrm{d}E' \  \frac{N(E')}{E'}, 
\end{align}
so that using Eqs.\ \eqref{eq:DoS_th} and \eqref{eq:DoS_th2}, 
\begin{align}
\xi(E) &= \begin{dcases}
\frac{1}{\sigma_u^2} |\ln E|, &\lambda > 2 \\
\frac{1}{b_2} \frac{|\ln E|}{\ln |\ln E|}, &\lambda = 2 \\
\frac{(\lambda-1) \tau_\lambda}{[\gamma(\lambda)]^\lambda} |\ln E|^{\lambda-1}, &1 < \lambda < 2. 
\end{dcases} \label{eq:xiThou}
\end{align}

For $\lambda \leq 1$, the Thouless relation suggests that $\xi(E) = 0$ since $\mu_u = \infty$. 
We conclude that the wave functions all decay faster than an exponential.

To verify these results using computational techniques, we use transfer matrices (Eq.\ \eqref{eq:TM}) to obtain wave functions on systems of size $N = 2 \times 10^7$ sites, at different energies.
We use the correlation function $g(r)$ as defined in Eq.\ \eqref{eq:gr} to extract the ensemble-averaged properties of the fall-off of the wave function.
In Fig.\ \ref{fig:xi_scaling2}, we plot the results for four representative values of $\lambda$.
At low energies and short length scales, the correlation function behaves as in the critical case at $E = 0$.
This situation is seen both in the universal Dyson class ($\lambda > 2$), as well as in the non-universal case ($\lambda < 2$), in accordance with the exact non-universal scaling results derived in Sec.\ \ref{sec:E0}.
At larger distances, $g(r)$ eventually becomes linear with $r$ when $\lambda > 1$.
When $\lambda < 1$, the entire spectrum is influenced by the super-exponential decay of the zero-energy state, and so $g(r)$ scales as $r^{1/\lambda}$.

\begin{figure}[ht!]
\centering
\includegraphics[angle=0,width=\columnwidth]{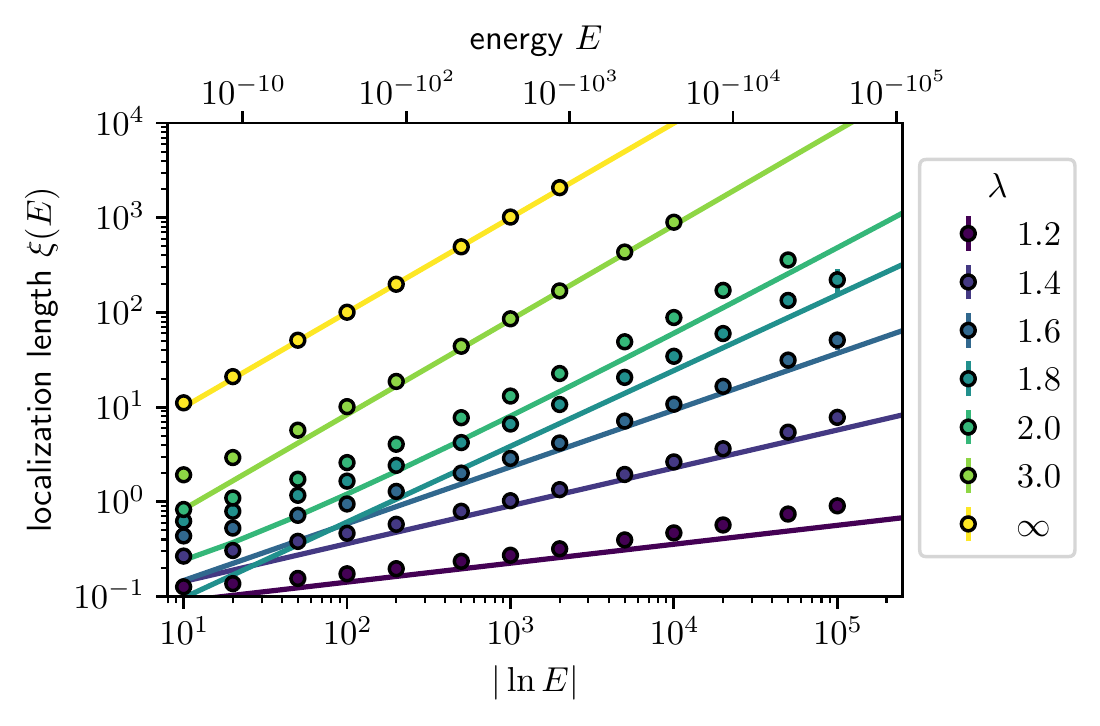} 
\caption{\label{fig:xi_scaling}
The localization length as a function of energy for different values of $\lambda$, both within and outside the Dyson class.
Circles are obtained numerically from the asymptotic value of the scaled correlation functions $g(r)$, e.g.\ those in Fig.\ \ref{fig:xi_scaling2}.
Solid lines are the theoretical prediction Eq.\ \eqref{eq:xiThou}.}
\end{figure}

These results are all as expected.
Further, from the scaling of the correlation function at large $r$ as a function of energy $E$, we can test the predictions in Eq.\ \eqref{eq:xiThou}.
In Fig.\ \ref{fig:xi_scaling}, we plot $\xi(E)$ versus $|\ln E|$, where the numerical values for localization length are obtained as \begin{align}
\xi(E) = \frac{1}{\lim\limits_{r \to \infty} g(r)/r}.
\end{align}
We find that within a regime of intermediate energies, the correct functional relationship as suggested by Eq.\ \eqref{eq:xiThou}, including the prefactors, is obtained.
When $E$ is relatively large, then higher-order terms reduce the accuracy, and when $E$ is very small, the wave function obtained is limited by the finite size of the system, and the thermodynamic limit is not seen.
This is a remarkable agreement between theory and numerics, covering both the Dyson class and the non-universal case $\lambda < 2$.

We conclude with a remark on another measure, the inverse participation ratio, often used to quantify localization in numerical studies of disordered systems.
We find that it does not yield any useful information here, due to the stretched exponential nature of the wave function envelope.
Nevertheless, for completeness, we provide a brief summary of our calculations of the inverse participation ratio in Appendix \ref{IPR}.

\section{Magnetic susceptibility of random spin chain models} \label{sec:exp}

The spinless fermionic nearest neighbor hopping model directly maps to a random nearest neighbor antiferromagnetic XY spin chain.
In that case, the primary experimental quantity of interest is the variation of magnetic susceptibility as a function of temperature.
The magnetic susceptibility $\chi(T)$ is related to the number of free spins with effective couplings below $T$.
The well-known universal behavior is\cite{Fisher1994} \begin{align}
\chi(T) \sim \frac{1}{T \ln^2 (\Omega/T)},
\end{align}
where $\Omega$ is an energy scale intrinsic to the system.

As a direct consequence of our results for the integrated density of states in Sec.\ \ref{sec:DoS} (Eqs.\ \eqref{eq:DoS_th} and \eqref{eq:DoS_th2}), the susceptibility formula gets modified to 
\begin{align}
\chi(T) \sim \begin{dcases}
\frac{1}{T \ln^{2}(\Omega/T)}, \quad \lambda > 2, \\
\frac{ \ln \ln (\Omega/T)}{T \ln^{2}(\Omega/T)}, \quad \lambda = 2, \\
\frac{1}{T \ln^{\lambda}(\Omega/T)}, \quad \lambda < 2.
\end{dcases}
\end{align}
when the antiferromagnetic couplings are chosen from a distribution of the form in Eq.\ \eqref{eq:p_lambda}, with a tunable divergence at zero.

While the Heisenberg spin chain does not map to a non-interacting fermionic model, following the arguments of Fisher\cite{Fisher1994}, we expect the same behavior to hold in the case of the random Heisenberg spin chain as well.

\section{Conclusions} \label{sec:concl}

\begin{table*}
\begin{tabularx}{\textwidth}{@{} X | p{1mm} X X X X X @{}}
\toprule
\toprule
  & & $\bm{0 < \lambda < 1}$ \bigstrut & $\bm{\lambda = 1}$ & $\bm{1 < \lambda < 2}$ & $\bm{\lambda = 2}$ & $\bm{\lambda > 2}$ (Dyson class)\\
\specialrule{.8pt}{2pt}{2pt}
\multicolumn{7}{c}{statistics of hoppings $t$ and their distribution $p_\lambda(t)$} \bigstrut \\
\midrule
  normalization $c_\lambda$ \bigstrut & & $\lambda(\lambda + 1)^\lambda$ & $2$ & $\lambda(\lambda + 1)^\lambda$ & $18$ & $\lambda(\lambda + 1)^\lambda$\\

  median $t$ \bigstrut & & 1 & 1 & 1 & 1 & 1\\
\specialrule{.8pt}{2pt}{2pt}
 
  \multicolumn{7}{c}{statistics of the logarithm of hoppings $u \equiv - \ln t$} \bigstrut \\
\midrule
    
  mean $\mu_u$ \bigstrut & & $\infty$ & $\infty$ & finite & finite & finite\\
 
  variance $\sigma_u^2$ \bigstrut & &
  $\infty$ & 
  $\infty$ & 
  $\infty$ & 
  $\infty$ & 
  {$\begin{aligned}\frac{\lambda (\lambda+1)^2}{(\lambda-2)(\lambda-1)^2}\end{aligned}$} \bigstrut \\ 
\specialrule{.8pt}{2pt}{2pt}
  \multicolumn{7}{c}{wave function of the band-center state ($E = 0$)} \bigstrut \\ 
\midrule
  $\psi(r)$ & &
  {$\begin{aligned}e^{-(r/r_o)^{1/\lambda}}\end{aligned}$}& 
  {$\begin{aligned}e^{-r/r_0}\end{aligned}$} & 
  {$\begin{aligned}e^{-(r/r_0)^{1/\lambda}}\end{aligned}$} & 
  {$\begin{aligned}e^{-\sqrt{(r/r_0) \ln (r/r_0)}}\end{aligned}$}  \bigstrut & 
  {$\begin{aligned}e^{-\sqrt{r/r_0}}\end{aligned}$}\\
decay envelope \bigstrut & & super-exponential & exponential & sub-exponential & sub-exponential & sub-exponential\\

%
  distribution of $\frac{1}{r_0}$ \footnote{in this row, $\mathcal{S}$ refers to the \emph{folded} symmetric alpha-stable distribution, and $\mathcal{N}$ to the folded Gaussian distribution, each with support only for positive values} & &
  {$\begin{aligned}\mathcal{S}(\lambda, 0, \gamma(\lambda), 0)\end{aligned}$} \footnote{the scale factor $\gamma(\lambda)$ is $\left( \frac{\pi c_\lambda}{2 \lambda \sin(\frac{\pi \lambda}{2}) \Gamma(\lambda)} \right)^{\frac{1}{\lambda}}$} & 
  {$\begin{aligned}\mathcal{S}(1, 0, \pi, 0)\end{aligned}$} & 
  {$\begin{aligned}\mathcal{S}(\lambda, 0, \gamma(\lambda), 0)\end{aligned}$} & 
  {$\begin{aligned}\mathcal{N} \left( 0, \frac{c_\lambda}{2} \right) \end{aligned}$} \bigstrut & 
  {$\begin{aligned}\mathcal{N}(0, \sigma_u^2)\end{aligned}$} \\
%
\specialrule{.8pt}{2pt}{2pt}
  \multicolumn{7}{c}{integrated density of states $N(E)$ as $E \to 0$} \bigstrut \\
\midrule
  $N(E)$ & & {$\begin{aligned}  \frac{b_\lambda}{|\ln E|^\lambda} \end{aligned}$} \bigstrut & 
  {$\begin{aligned}\frac{b_\lambda}{|\ln E|}\end{aligned}$} & 
  {$\begin{aligned}\frac{b_\lambda}{|\ln E|^\lambda}\end{aligned}$} & 
  {$\begin{aligned}\frac{b_\lambda \ln |\ln E| }{|\ln E|^2}\end{aligned}$} & 
  {$\begin{aligned}\frac{b_\lambda}{|\ln E|^2}\end{aligned}$} \\
  pre-factor $b_\lambda$  \bigstrut & &
  {$\begin{aligned}\frac{[\gamma(\lambda)]^\lambda}{\tau_\lambda}\end{aligned}$} \footnote{see Fig.\ \ref{fig:tau} in Appendix \ref{apptau} for details on the mean first passage time $\tau_\lambda$} & 
  {$\begin{aligned}2\end{aligned}$} & 
  {$\begin{aligned}\frac{[\gamma(\lambda)]^\lambda}{\tau_\lambda}\end{aligned}$} \bigstrut & 
  {$\begin{aligned}18\pm 2\end{aligned}$} & 
  {$\begin{aligned}\sigma_u^2\end{aligned}$} \\
\specialrule{.8pt}{2pt}{2pt}
  \multicolumn{7}{c}{spatial profile and localization length $\xi(E)$ at non-zero energy} \bigstrut \\
\midrule
decay envelope \bigstrut & & super-exponential & super-exponential & exponential & exponential & exponential\\

 $\xi(E)$ & &
 $0$ & 
 $0$ & 
 {$\begin{aligned}\frac{(\lambda-1) \tau_\lambda}{[\gamma(\lambda)]^\lambda} |\ln E|^{\lambda-1}\end{aligned}$} \bigstrut &
 {$\begin{aligned}\frac{1}{b_2} \frac{|\ln E|}{\ln |\ln E|}\end{aligned}$} & 
 {$\begin{aligned}\sigma_u^2 |\ln E|\end{aligned}$} \\
\bottomrule
\bottomrule
\end{tabularx}
\caption{Summary of key findings for the one-dimensional chain with random hopping studied in the paper.
The hopping distribution $p_\lambda(t)$ (Eq.\ \eqref{eq:p_lambda}) is parametrized by $\lambda$.
For $\lambda > 2$, we recover the universal `Dyson' behavior, but for $\lambda \leq 2$, there is a variety of non-trivial physics.
\label{tab:table1}
}
\end{table*}

In this paper, we have studied a non-interacting spinless fermionic system with nearest neighbor hopping (off-diagonal) disorder, using a combination of analytic methods, numerical strong disorder renormalization group and transfer matrix techniques.
This model owes its peculiar behavior in large part to the exact sub-lattice symmetry.
The salient feature of our work is that the disordered hopping matrix elements $t$ are drawn from a probability density with a tunable sharp divergence as $t \to 0$.
The divergence is of the form $p(t) \sim 1/t \ln^{\lambda+1}(1/t)$.
While we have written our results in the language of an electronic model, the conclusions are directly applicable to the magnetic properties of spin chains, as summarized in Sec.\ \ref{sec:exp}.
While this work was being completed, we discovered that in his seminal work on random antiferromagnetic spin chains\cite{Fisher1994}, Daniel Fisher did point out that if the couplings $J$ are distributed as $p(J) \sim 1/(J |\ln J|^x)$, then the properties of the system are dominated by weak links, and the behavior is expected to be non-universal. 
However, there does not appear to have been a thorough investigation of the issue until now.

In this work, we find that when $\lambda > 2$, we recover the well known universal behavior of the Dyson class.
In all such cases the density of states diverges as $\rho(E) \sim 1/(E |\ln E|^3)$ and the localization length diverges as $\xi(E) \sim |\ln E|$ near the critical point $E = 0$.
The wave function at zero energy is not exponentially localized, and falls off as $\exp\left( - \sqrt{r/r_0}\right)$.

As we show here, non-universality is only seen when the exponent $\lambda < 2$.
There is a whole line of fixed-points $0 < \lambda < 2$ that lies outside the basin of attraction of the standard Dyson model.
In this case, the state at zero-energy decays as $\exp \left( - (r/r_0)^{1/\lambda} \right)$, which is super-exponential for $\lambda < 1$.
The density of states and localization length also demonstrate non-universal and continuously tunable behavior.
At the boundary of the Dyson class, $\lambda = 2$, the wave function envelope, density of states and localization lengths show non-trivial logarithmic divergence.
By systematically exploring the whole parameter regime, we have established the existence of non-universal behavior and characterized its nature.
Table \ref{tab:table1} summarizes the central results of our paper.

We emphasize that in this model there are extremely large variations in the hopping terms, over several orders of magnitude.
This is evidenced by several of the plots on this paper being on an iterated logarithmic scale in order to capture the full dynamic range, often approaching energy values of the order of $10^{-10^8}$.
We took extra care in the numerics, in order to avoid floating point overflow and underflow errors.
From an experimental point of view, it might seem that such hoppings are challenging to realize in practice.
Nevertheless, as we show in Appendix \ref{app1}, it may be possible to construct such nearest neighbor couplings in synthetic lattices, using a modified form of the Morse potential.

The non-trivial behavior of our model is entirely due to our choice of a probability distribution with fat tails.
This is similar to the origin of non-universal anomalous diffusion seen in L\'{e}vy flights and other kinds of non-Brownian stochastic processes.
While we have made a mapping between the low-energy properties of a quantum Hamiltonian and a first passage problem in classical non-equilibrium statistical mechanics, there are surely deeper connections waiting to be explored. 
Anomalous diffusion and L\'{e}vy flights are currently topics of active research, with ongoing efforts to push the boundaries of both mathematical theory and applications to other fields, including chemical kinetics, evolutionary biology, economics and finance.
In this context, it would be of interest to examine other quantum condensed matter systems where extreme value statistics and effects of large deviations cause a breakdown of universal behavior.
Implications of this to systems in dimensions greater than one, or random graphs, \emph{maintaining the bipartite nature of the problem}, are also left for future work.

\acknowledgments

This project was conceived while R.N.B. was at the Aspen Center for Physics, which is supported by the NSF.
It was completed during R.N.B.'s sabbatical stay at IAS, just prior to the passing away of Freeman Dyson, an IAS icon for several decades.
R.N.B. acknowledges brief discussions with Kedar Damle, David Huse, Gil Refael and Thomas Spencer.
A.K. thanks Gerasimos Angelatos for useful discussions.
We acknowledge support from Department of Energy BES grant DE-SC0002140.

\appendix

\section{Mean first passage time for the L\'{e}vy flight} \label{apptau}

\begin{figure}[ht!]
\centering
\includegraphics[angle=0,width=\columnwidth]{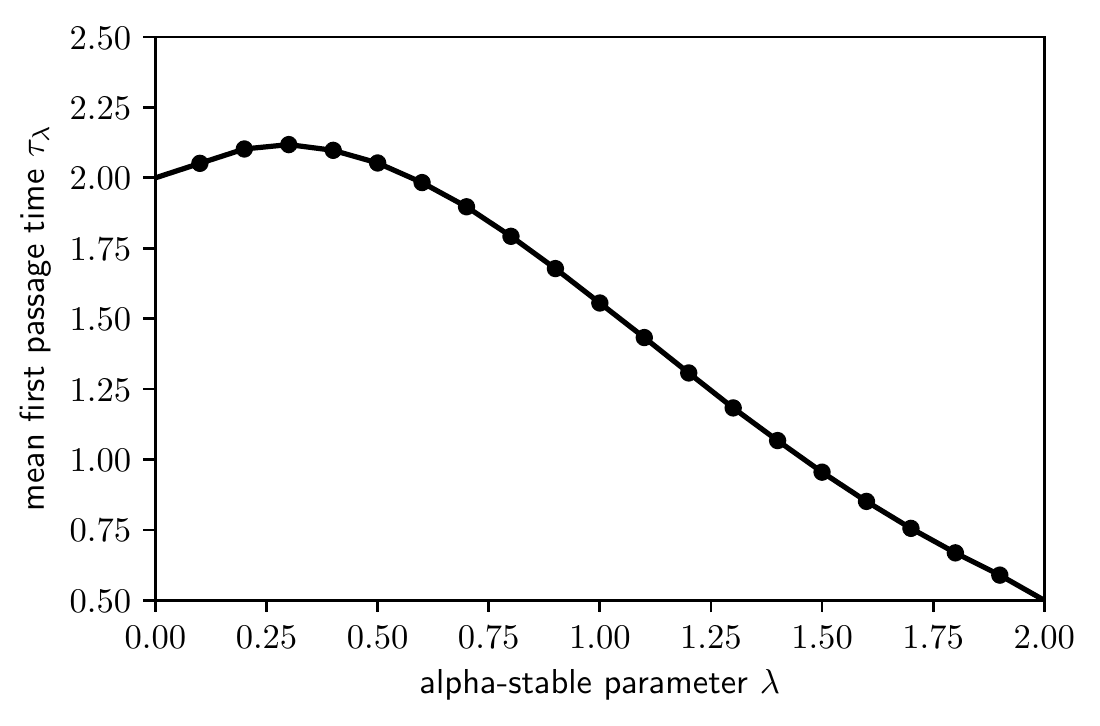} 
\caption{\label{fig:tau}
The mean first passage time $\tau_\lambda$ calculated by numerical simulations.}
\end{figure}

In this Appendix, we closely follow Ref.\ \cite{Dybiec2017}.
The normalized L\'{e}vy flight is given by the stochastic process \begin{align}
\frac{\mathrm{d}x}{\mathrm{d}t} = \zeta_\lambda(t),
\end{align}
which is discretized to \begin{align}
\Delta x = (\Delta t)^{1/\lambda} \zeta_\lambda,
\end{align}
where $\zeta_\lambda$ is a random variable drawn from the symmetric alpha-stable distribution $\mathcal{S}(\lambda, 0, 1, 0)$.
The particle moves in a domain with an infinite hard wall $V(x) = \infty$ for $x<0$.
The hard wall acts like a stopping boundary: if any increment were to send the particle to $x<0$, it stops at $x=0$.
Then the mean first passage time is
\begin{align}
\tau_\lambda = \langle \min \{ t>0 : x(0) = 0 \text{ and } x(t) > 1 \} \rangle,
\end{align}
where the angular brackets denote an ensemble average.
We use the $\Delta t = 10^{-4}$ and average over $N=10^5$ trajectories for each value of $\lambda$ to obtain the points plotted in Fig.\ \ref{fig:tau}.
These numbers are then used in Eq.\ \eqref{eq:DoS_th} for the integrated density of states to obtain the curve in Fig.\ \ref{fig:DoS1}.

\section{Inverse Participation Ratios} \label{IPR}

The inverse participation ratio (IPR) is a commonly used measure of localization, defined in terms of wave function amplitudes as \begin{align}
\mathsf{IPR} \equiv \frac{\sum_i |\psi_i|^4}{\left( \sum_i |\psi_i|^2 \right)^2}.
\end{align}

For exponentially localized (e.g.\ Anderson insulator) states, the IPR is relatively large and constant with size, while for extended (metallic states), the IPR scales as $\sim 1/L^d$, where $L$ is the linear extent of the system and $d$ is its dimension.
For critical states (e.g.\ at the Anderson metal-insulator transition or center of a Landau level), the IPR and its moments scale anomalously with size.
They contain useful information about the multifractal nature of the critical state and the tails of its wave function envelope, which often have power-law tails.

\begin{figure}[ht!]
\centering
\includegraphics[angle=0,width=\columnwidth]{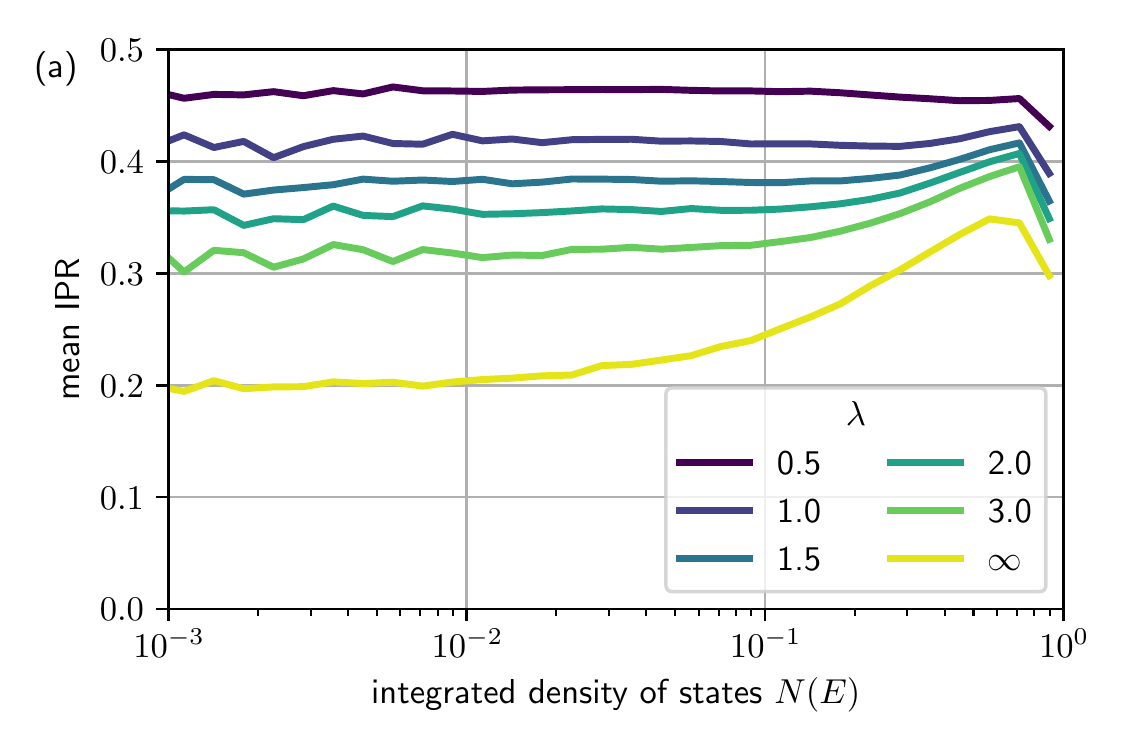}
\includegraphics[angle=0,width=\columnwidth]{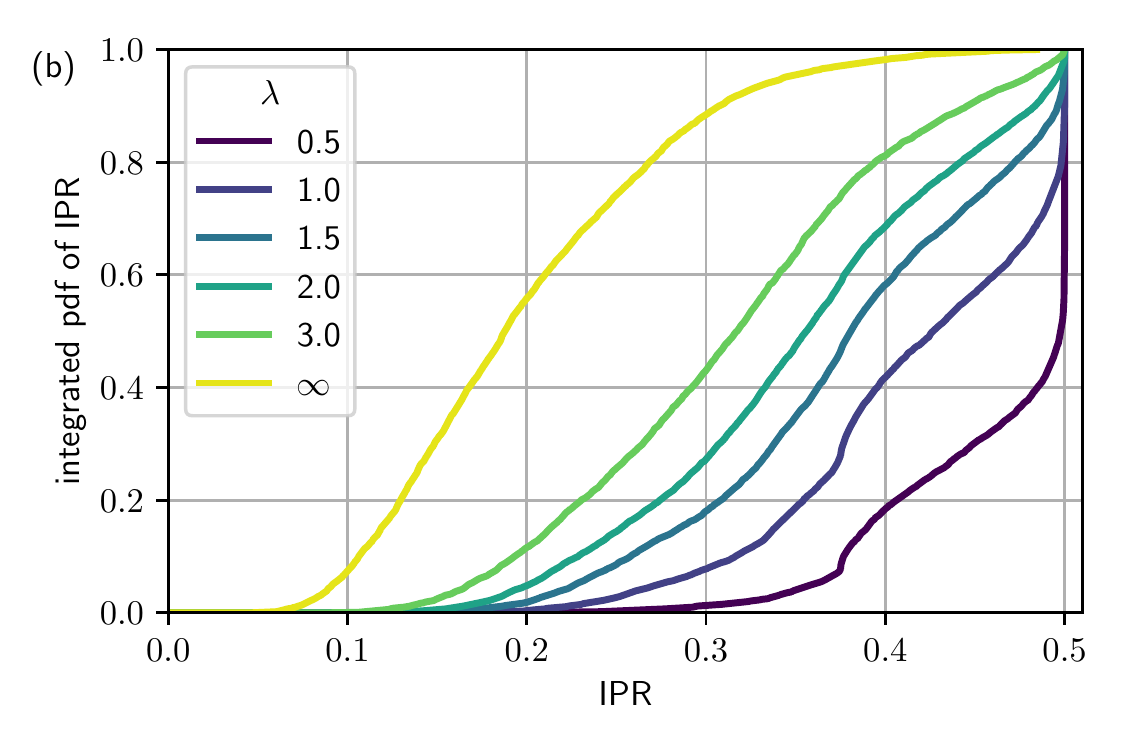} 
\caption{\label{fig:IPR}
(a) The energy-resolved mean inverse participation ratio as a function of the integrated density of states $N(E)$ is largely featureless.
These results are obtained by ensemble averaging approximately 1000 realizations of disorder on systems of size $N=10^4$.
(b) The integrated probability density function (pdf) of the IPR for states with the lowest $1\%$ of energies.
There is zero weight for small IPR, indicating the complete absence of conventionally delocalized states over the full range $0 < \lambda < \infty$.}
\end{figure}

However, in this model, both in the Dyson class ($\lambda > 2$) and beyond ($\lambda \leq 2$), we find that the IPR is not a useful metric, even though there is a critical point and the wave functions are not conventionally localized.
This seemingly counter-intuitive situation is because the wave functions have stretched exponential tails, whose signature cannot be captured by the IPR.
Hence, other diagnostics, such as the Lyapunov exponent or the correlation function (Eq.\ \eqref{eq:gr}) must be used to quantify localization.
We illustrate this fact in Fig.\ \ref{fig:IPR} (a), where we provide mean IPRs as a function of energy for a variety of $\lambda$.

The IPR is useful in extracting a length scale $\xi_{\text{IPR}} \sim \mathsf{IPR}^{-1}$, which is the average number of sites on which the state has most of its weight.
Here, it seems apparent that the wave functions are all localized on an average of 5 sites or fewer, with the degree of localization increasing as $\lambda$ decreases.
However, the IPR gives us no information about the decay of the wave function envelope.
We also examined the distribution of IPRs (see Fig.\ \ref{fig:IPR} (b) for a typical case).
As can be seen, the probability distribution of IPR decays extremely rapidly for IPR $\to 0$ (i.e.\ $\xi_{\text{IPR}} \to \infty$), so there do not appear to be any signatures of the broad distributions of the decay length $r_0$ described in Sec.\ \ref{sec:xi}.

\section{Tailoring the hopping distribution in experiment} \label{app1}

It may appear that the singular distributions $p(t)$ considered in this paper are not possible in practice, since wave functions $\psi(r)$ generically decay exponentially with $r$, without fine tuning.
This leads to hoppings $t(r)$ that also decay exponentially with $r$.
It would then require setting up a chain with nearest neighbor spacings that are rather contrived to obtain a $p(t)$ that is of the desired form.
However, in this age, with atoms in an engineered optical lattice, it is more likely that one can configure a random chain with nearest neighbor spacings from a more realistic distribution, but instead engineering the wave function to decay differently with $r$.
We describe one such possibility in this section.

Consider a chain of atoms in which the positions are purely randomly distributed (i.e.\ Poisson statistics). In such a chain, the probability of not having any neighbor within a distance $R$ (on either side) goes as:
\begin{align}
N(R) = \exp ( - 2 \rho R)
\end{align}
where $\rho$ is the number density of sites. This leads to the nearest neighbor probability density function as:
\begin{align}
\phi(R) = \frac{dN}{dR} = 2 \rho \exp (- 2 \rho R)
\end{align}

Now, if the nearest neighbor hopping goes as $t(R)$, which we assume to be a monotonically decreasing function of $R$, the probability distribution function of $t$ will be given by
\begin{align}
p(t) = \phi(R) \frac{dR}{dt}
\end{align}
and demanding $p(t) =  \frac{c_\lambda}{|t \ln^{\lambda + 1}(1/t)|}$ immediately yields
\begin{align}
t(R) \sim \exp(- C  e^{\frac{\rho R}{\lambda}})
\end{align}
where $C = (\frac{c_\lambda}{2 \lambda})^{\frac{1}{\lambda}}$.

The above requires, within a tight-binding approximation where $t(R) \simeq \int \mathrm{d}r \ \psi(r) \psi^\ast(R-r)$, an on-site wave function $\psi(r)$ whose decay also has a similar functional form (plus sub-leading factors)
\begin{align}
\psi(r) \sim \exp( - c e^{b r} ), 
\end{align}
with $c = C/2$, and $b = \frac{2 \rho}{\lambda}$. 
Putting that in the single site Schrodinger equation and demanding that it be an eigenfunction for a particle of mass $m$ with energy $E$ implies that the potential must have inversion symmetry $V(r) = V(-r)$ and be of the form: 
\begin{align}
V (r) = E + \frac{\hbar^2}{2m} b^2 c \left( c e^{2br} - e^{br} \right), \ \text{for } r > 0. \label{eq:expPot}
\end{align}

The above result is functionally similar to the Morse potential\cite{Morse1929} used extensively in studying vibrations of diatomic molecules, where it takes the form:
\begin{align}
V_M(r) = V_0 \left( e^{-2a(r-r_0)} - 2e^{-a(r-r_0)} \right) \label{eq:Morse}
\end{align}

However, unlike the standard Morse potential, which exhibits a minimum value $-V_0$ at $r = r_0$, the potential in Eq.\ \eqref{eq:Morse} has no such minimum for parameter values of interest here (positive $b$ and $r$), so we refer to it as the \emph{generalized} Morse potential.

Engineering an exponential potential of the form \eqref{eq:expPot} is a challenge we leave to the ingenuity of practitioners of AMO lattices. 

\bibliography{Dyson_papers}

\end{document}